\documentclass[a4paper,11pt]{article}
\pdfoutput=1 

\usepackage{jinstpub} 

\title{Test and characterization of 400 Hamamatsu R5912-MOD photomultiplier
tubes for the ICARUS T600 detector}

\collaboration{The ICARUS/NP01 Collaboration}

\author[a,b]{M.~Babicz,}
\author[c]{L.~Bagby,}
\author[d]{B.~Baibussinov,}
\author[e]{V.~Bellini,}
\author[f]{M.~Bonesini,} 
\author[d,g]{A.~Braggiotti,} 
\author[d]{S.~Centro,}
\author[h]{T.~Cervi,}
\author[i]{A.G.~Cocco,}
\author[h,l]{A.~Falcone,} 
\author[d]{C.~Farnese,}
\author[c,d]{A.~Fava,}
\author[e]{F.~Fichera,}
\author[d]{D.~Gibin,}
\author[d]{A.~Guglielmi,}
\author[b]{U.~Kose,}
\author[f]{R.~Mazza,}
\author[h]{A.~Menegolli,}
\author[d]{G.~Meng,}
\author[h]{C.~Montanari,}
\author[b]{M.~Nessi,}
\author[h]{P.~Picchi,}
\author[b,d]{F.~Pietropaolo,}
\author[h]{M.C.~Prata,}
\author[h]{A.~Rappoldi,}
\author[h,1]{G.L.~Raselli\note{Corresponding author.},}
\author[h]{M.~Rossella,}
\author[b,m,n,2]{C.~Rubbia\note{Spokesman.},} 
\author[b]{P.~Sala,}
\author[h]{A.~Scaramelli,}
\author[o]{F.~Sergiampietri,}
\author[h,3]{M.~Spanu\note{Now at Brookhaven National Laboratory, NY, USA.},}
\author[f]{M.~Torti,}
\author[e]{F.~Tortorici,}
\author[d]{F.~Varanini,}
\author[d]{S.~Ventura,}
\author[n]{C.~Vignoli}
\author[b]{and A.~Zani}

\affiliation[a]{Institute of Nuclear Physics PAN, Cracow, Poland}
\affiliation[b]{CERN, Geneva, Switzerland}
\affiliation[c]{Fermi National Laboratory, Batavia IL, USA}
\affiliation[d]{University of Padova and INFN, Padova, Italy}
\affiliation[e]{University of Catania and INFN, Catania, Italy}
\affiliation[f]{University of Milano Bicocca and INFN, Milan, Italy}
\affiliation[g]{CNR, Padova, Italy}
\affiliation[g]{University of Pavia and INFN, Pavia, Italy}
\affiliation[i]{University of Napoli and INFN, Napoli, Italy}
\affiliation[l]{University of Texas in Arlington, Arlington TX, USA}
\affiliation[m]{Gran Sasso Science Institute, L'Aquila, Italy}
\affiliation[n]{INFN, Laboratori Nazionali del Gran Sasso, Assergi, Italy}
\affiliation[o]{INFN, Pisa, Italy}

\emailAdd{gianluca.raselli@pv.infn.it}

\abstract {
ICARUS T600 will be operated as far detector of the Short Baseline Neutrino program at Fermilab (USA), which foresees
three liquid argon time projection chambers along the Booster Neutrino Beam line to search for a LSND-like sterile
neutrino signal.
The detector employs 360 photomultiplier tubes, Hamamatsu model R5912-MOD, 
suitable for cryogenic applications.
A total of 400 PMTs were procured from Hamamatsu
and tested at room temperature to evaluate the performance of the devices and their
compliance to detect the liquid argon scintillation light in the T600 detector.
Furthermore 60 units were also characterized at cryogenic temperature, 
in liquid argon bath, to evaluate any parameter variation which could affect the scintillation light detection.
All the tested PMTs were found to comply with the requirements of ICARUS T600
and a subset of 360 specimens was selected for the final installation in the detector.}


\begin{document}
\maketitle


\section{Introduction}

ICARUS T600 detector is the largest Liquid Argon Time Projection Chamber (LAr-TPC)
ever built for neutrino oscillation studies, successfully 
operated at LNGS (Laboratori Nazionali del Gran Sasso) underground laboratory (Italy)~\cite{T600}.
It took data from 2010 to 2013
exposed to the CNGS (CERN Neutrinos to Gran Sasso) beam from CERN and atmospheric 
neutrinos with a live-time factor in excess of 93\%~\cite{LNGS}.
After an intense refurbishing operation, carried out at CERN
in the framework of the Neutrino Platform
(NP01) activities, 
the entire apparatus was transferred to Fermilab (IL, USA). It will become the far detector 
of the Short Baseline Neutrino (SBN) program~\cite{SBN}, exploiting three liquid argon detectors, placed along the Booster Neutrino Beam (BNB) line and operating at shallow depth, to investigate the possible presence of sterile neutrino states.

The realization of a new liquid argon scintillation light detection system, with high performance photomultiplier tubes (PMTs) represents a primary task 
of the detector refurbishing. Since the detector  
will be subject to a huge cosmic flux, the light detection system would allow the 3D reconstruction of events
contributing to the identification of neutrino interactions in the beam spill gate.

The T600 scintillation light detection system was significantly upgraded
at CERN from summer 2015 to summer 2017. The 74 outdated ETL 9357FLA PMTs mounted in the detector 
in 2000~\cite{T600,ETL} were replaced with new 360 Hamamatsu R5912-MOD units operating in liquid argon.
Each PMT was tested and characterized 
to determine its basic functionality and to verify its compliance with the required functioning specifications. 
In particular all the PMTs were tested at room temperature and 60 units were also characterized at cryogenic temperature, in liquid argon bath.
Measurements were carried out in different areas at CERN, where a dedicated dark-room and a cryogenic test facility were set-up. 

This paper describes this test campaign.
In particular, in section 2 the ICARUS T600 scintillation light detection system is described.
The main features of the R5912-MOD are presented in section 3. Section 4 describes the employed measurement
equipment and the CERN test facilities. Results are reported in section 5. A description of the base layout 
is also presented in Appendix~\ref{base}.

\section{The ICARUS T600 scintillation light detection system}

ICARUS T600 detector is made of two identical cryostats, each housing two TPCs with a
common central cathode. Charged particles interacting in liquid argon produce both scintillation
light and ionization electrons. These latter are drifted by a uniform electric field ($E=500$~V/cm)
to the anode, made with three parallel wire planes at 1.5~m from the cathode, where they are collected.
Photons are detected by a PMT system with the devices mounted
behind the wire planes and operating immersed in liquid argon (T = 87~K).
The information coming from both scintillation light and collected electrons 
allows obtaining a full 3D reconstruction of the interacting particle path.

\begin{figure}[htbp]
\centering
\includegraphics[width=0.8\columnwidth]{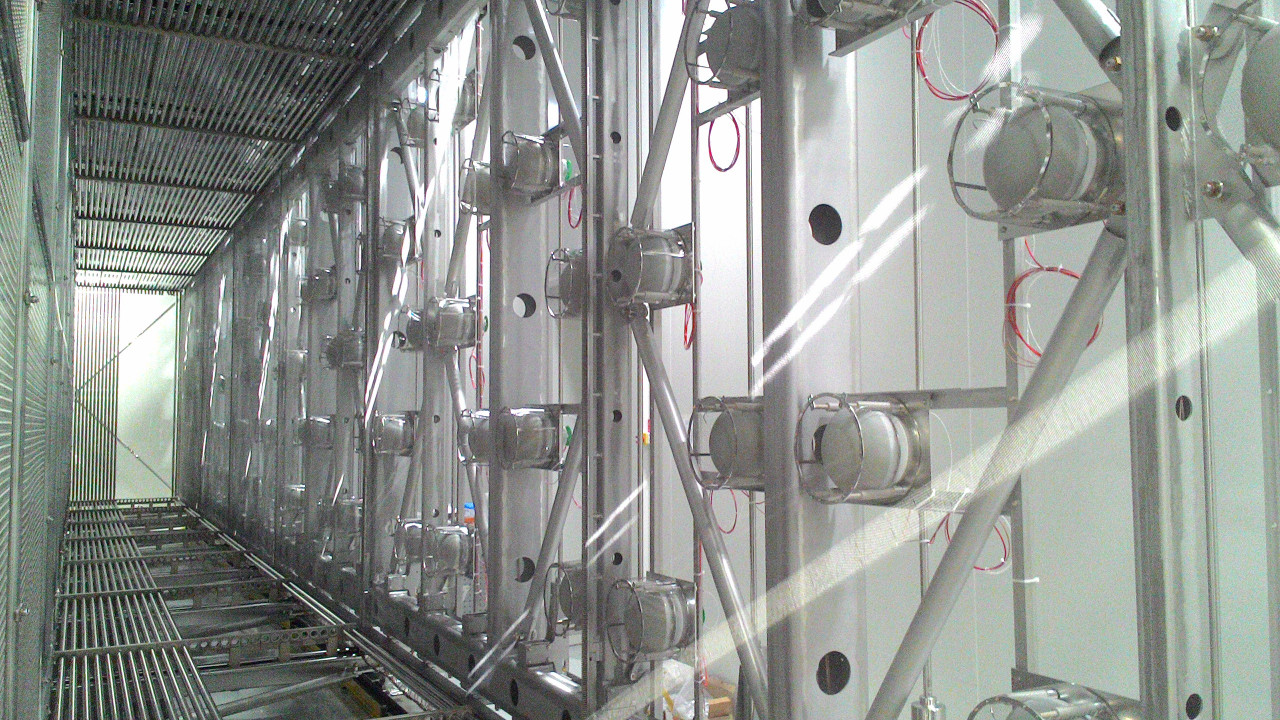}
\caption{PMT deployment behind a TPC wire plane.}
\label{fig_layout}
\end{figure}

The upgraded T600 light detection system consists of 360  Hamamatsu R5912-MOD PMTs 
deployed behind the 4 TPC wire chambers, 90 units each. 
Before installation the sensitive window of each PMT was coated with about 200$~\mu$g/cm$^2$ of 
Tetra-Phenyl Butadiene (TPB), a wavelength shifter to convert VUV LAr
scintillation photons ($\lambda = 128$~nm) to visible light. This process was performed using 
a dedicated thermal evaporator and a specific deposition technique~\cite{JVST,EVAP}.
Mechanical supports
were designed to hold up each PMT in the proper position at few millimeters 
behind the TPC wire planes. Each device is set inside a grounded stainless-steel 
cage to prevent the induction of PMT pulses on the facing wire plane. 
Furthermore a 50~$\mu$m optical fiber directed
towards the sensitive surface will allow a timing calibration at nanosecond precision.  
A picture of the new light detection system is shown in figure~\ref{fig_layout}.
 
\section{The Hamamatsu R5912-MOD}

\begin{figure}[!t]
\centering
\includegraphics[width=0.7\columnwidth]{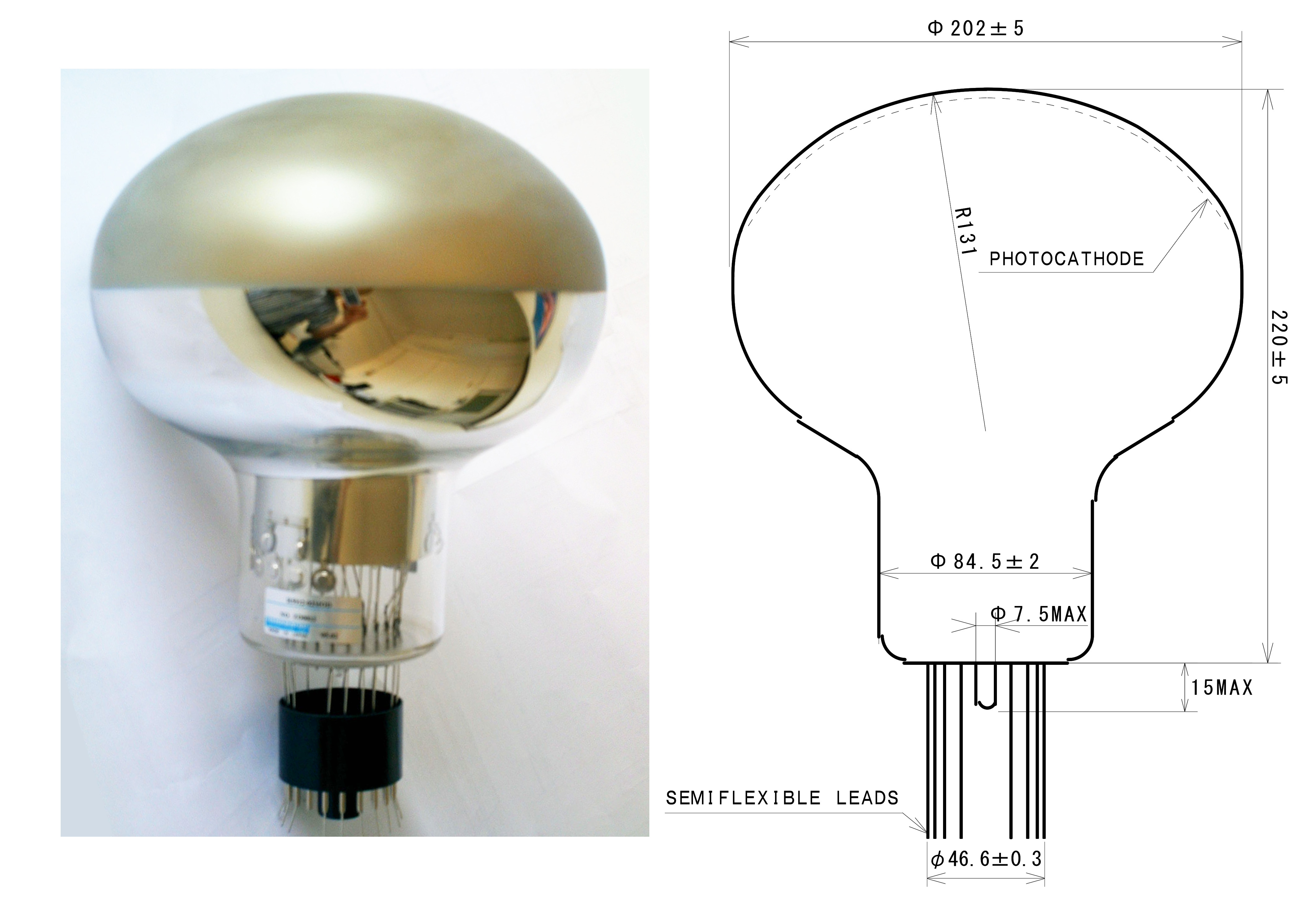}
\caption{The PMT Hamamatsu R5912-MOD. 
Lengths are in millimeters in the right figure~\cite{Hamamatsu}.}
\label{fig_pmt}
\end{figure}

The Hamamatsu R5912-MOD is a 10-dynode-stage PMT with an 8 in. hemispherical photocathode
(see figure~\ref{fig_pmt}).
The PMT is derived from the parent model R5912. The bialkali photocathode is deposited
on a Pt under-layer in order to extend the range of the operating temperature down to about -200~$^\circ$C.
This PMT model was selected  as a result of experimental tests carried out on different detectors
samples provided by different producers, such as Hamamatsu and ETL.
The main features of the Hamamatsu R5912-MOD resulting from these tests
can be summarized as follows~\cite{Falcone,Agnes,Timing}:
\begin{itemize}
\item Nominal PMT gain ($G\approx 10^7$) and quantum efficiency ($\approx 15\%$) which are
suitable for the detection of scintillation light produced by ionizing particle with energy $\le 100$~MeV; 
\item Good uniformity of the cathodic sensitivity all over the whole PMT window surface ($\pm 10$\%);
\item Good performances in terms of electron transit time spread ($\sigma \approx 1$~ns)
and electron transit time variation over the whole PMT window surface, both at room and in cryogenic environment;
\item Linear anodic response as a function of the incident light intensity,
both at room temperature and in cryogenic environment;
\item Pulse gain saturation above 300 photoelectrons.
\end{itemize}

\begin{table}[!t]
\renewcommand{\arraystretch}{1}
\caption{Main acceptance requirements for the Hamamatsu R5912-MOD}
\label{table1}
\centering
\begin{tabular}{ll}
\hline
Spectral Response      & $300\div 650$~nm \\
Window Material        & Borosilicate glass (sand blasted) \\
Photocathode           & Bialkali with Pt under-layer   \\
Max supply voltage (anode-cathode) & 2000 V \\
Photocathode Q.E. at 420nm  & $\ge 16$\%  \\
Typical Gain           &  $1\times 10^7$ at 1500 V\\
Nominal anode pulse rise time$^*$ & $\le 4$~ns\\
Nominal P/V ratio$^*$ & 2.5 \\
Max. dark count rate$^*$   & 5000~s$^{-1}$ \\
Max. transit time variation   & 2.5~ns (center-border)\\
\hline
$^*$ Values for $G =1\times 10^7$ & \\
\end{tabular}
\end{table}

A pre-series of 20 samples was delivered by Hamamatsu in 2015 for compliance testing purposes and
380 units in 2016.
PMTs were selected by the producer to fulfill the performance of table~\ref{table1}
at room temperature, only. All the PMTs were immersed in liquid nitrogen in factory
to certify their mechanical strength at cryogenic temperature.

Each device was provided with a proper base electronic circuit designed
according to the voltage distribution
ratio recommended by Hamamatsu.
This supplies the high voltage for the grids, dynodes and anode, and allows 
the signal to be picked up directly from the anode. Design and specifications are presented
in Appendix~\ref{base}.
Bases were directly welded on the
PMT flying leads. Since a negative power supply is adopted,
two independent coaxial cables are used to provide 
the PMT with the high voltage and
to read out the anode signal.

\section{Measuring equipment}

\label{sec_equip}

All the 400 samples were tested at room temperature and 60 of them were also qualified
at cryogenic temperature, in liquid argon bath. Characterization 
processes and measurement methods were all discussed and agreed with the producer. 

\begin{figure}[!t]
\centering
\includegraphics[width=0.4\columnwidth]{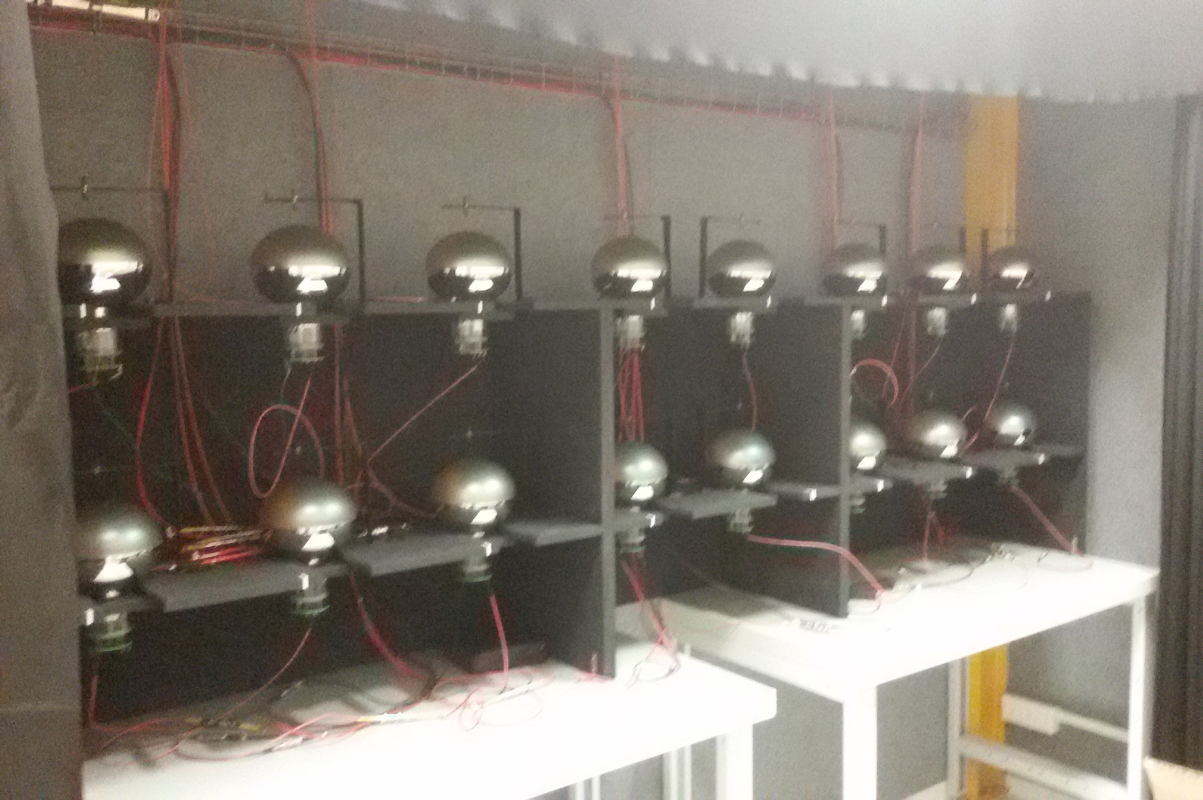}\hspace{0.1\columnwidth}
\includegraphics[width=0.4\columnwidth]{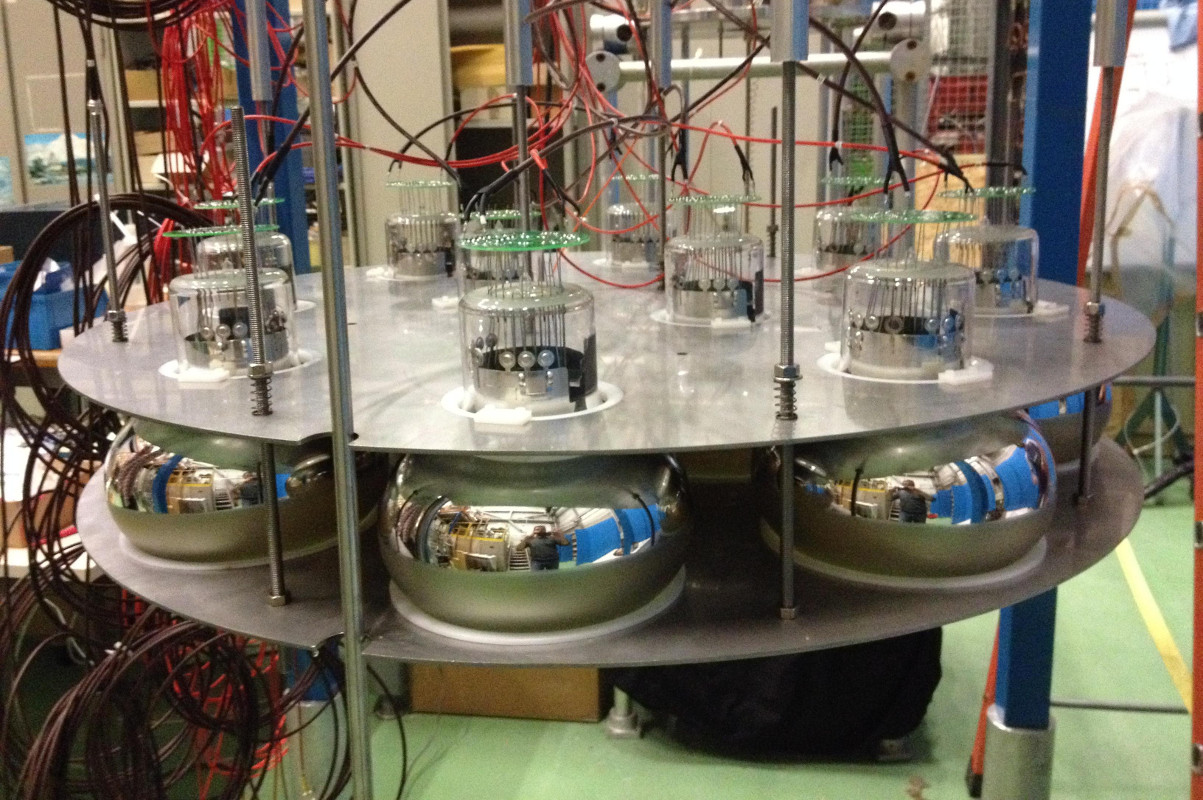}
\caption{CERN test facilities for room (left) and cryogenic temperature (right)
measurement.}
\label{fig_cern}
\end{figure}

\begin{figure}[!t]
\centering
\includegraphics[width=0.6\columnwidth]{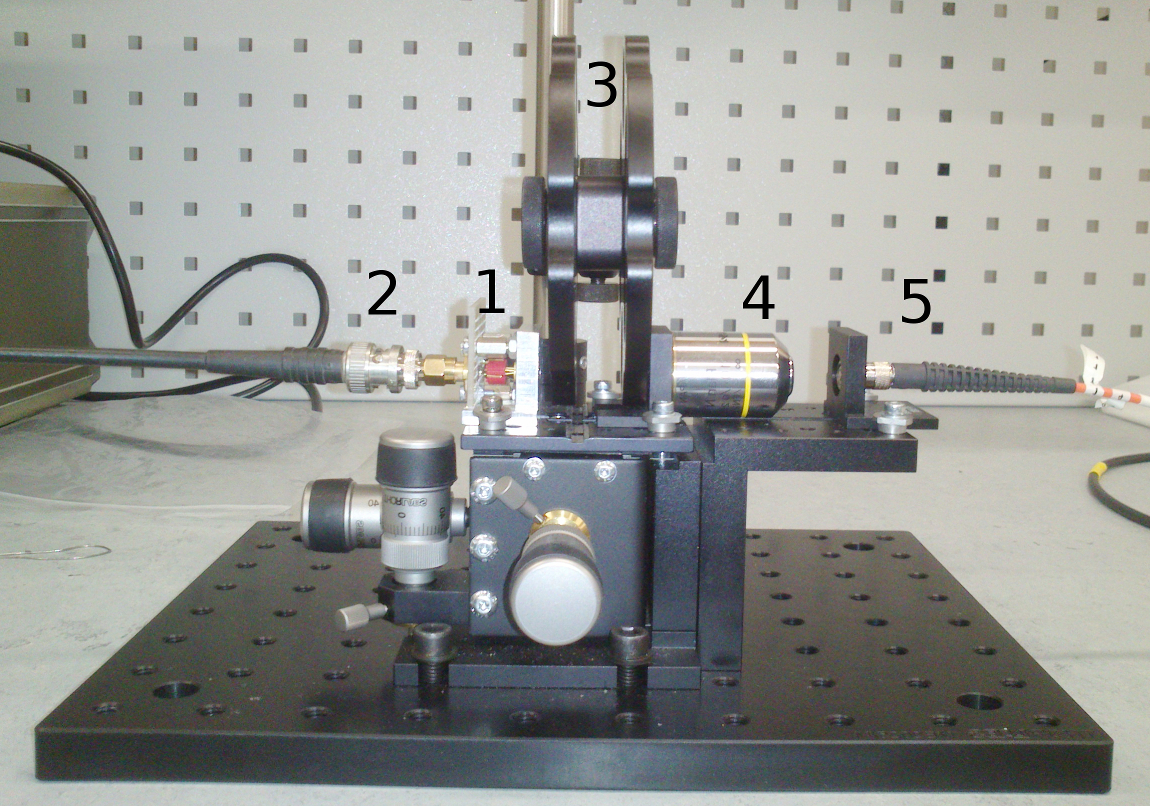}
\caption{Picture of the adopted laser instrumentation: 1) laser diode; 2) cable from
the pulser; 3) rotating filter supports; 4) focusing lens;
5) optical fiber towards the PMT under test.}
\label{fig_laser}
\end{figure}

A dark-room was arranged at CERN 
to test in parallel up to 16 PMTs (figure~\ref{fig_cern} left).
In a contiguous electronic workshop,
a laser diode was set-up to produce fast light pulses at 405~nm and about $1$~kHz repetition rate.
The light intensity was set through calibrated optical filters mounted on two-wheel supports, 
allowing different attenuation combinations ($1\div1000$). 
The light was focused on the PMT windows by means of 100~$\mu$m multi-mode optical fibers.
A picture of the instrumentation is shown in figure~\ref{fig_laser}.

\begin{figure}[!t]
\centering
\includegraphics[width=0.7\columnwidth]{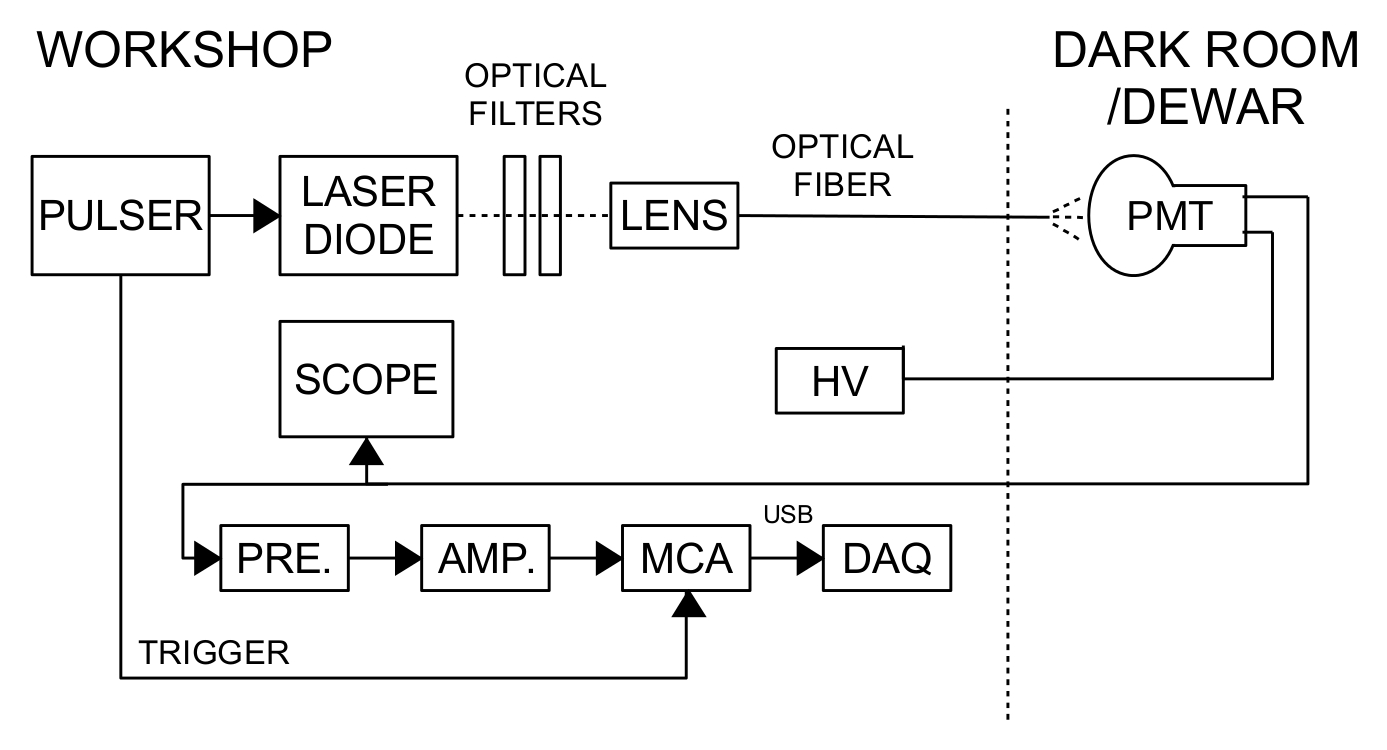}
\caption{Electronic set-up.}
\label{fig_setup}
\end{figure}

The characterization at room temperature  was carried out using
the electronics set-up shown in figure~\ref{fig_setup}.
The PMT output
was integrated by means of a CANBERRA 2005 charge
preamplifier and shaped by means of an ORTEC 570
amplifier. The output distribution was recorded by means
of a 8-k multichannel analyzer CAEN N915. Besides, PMT pulse shapes were acquired by means of  
a 10 GS/s oscilloscope (LeCroyWaveRunner 104MXI). 

Tests at cryogenic temperature were carried out in the building 182 at CERN. 
In order to obtain experimental conditions similar to those in the real apparatus,
PMTs were directly immersed in liquid argon (T = 87~K) inside a large dewar,,
allowing the simultaneous bath of 10 PMTs (figure~\ref{fig_cern} right).
The internal illumination was achieved by means of a
single 100~$\mu$m multi-mode optical fiber.
The same set-up and acquisition system described above was used
together with a proper feed-through, to preserve darkness conditions and thermal insulation.

\section{Tests and measurements}

Measurements at room temperature were carried out to evaluate the performance
of the devices and their
conformity to the requested features. 
Measurements were repeated on a sample of 60 PMTs at cryogenic temperature
to evaluate any parameter variation which could affect the scintillation light detection.

\subsection{Signal shape}

The shapes of the PMT anode pulses were recorded 
by means of a 10~GHz bandwidth oscilloscope. 
The PMT illumination was set in order
to have single photoelectrons emitted from the cathode,
while the resulting anode pulses where directly sampled
by the oscilloscope (50~$\Omega$ input impedance).
Each PMT under test was operated at different temperatures 
and at a multiplier gain
$G\approx 10^7$.
No significant variations were observed among the different tested samples
and the same mean values resulted
at room and at cryogenic temperature:
a leading edge of $3.9 \pm 1.1$~ns, a FWHM of $5.6 \pm 1.1$~ns and a
trailing edge of $10.3 \pm 1.6$~ns, in good agreement with the
nominal values indicated by the manufacturer. 

\subsection{Single electron response and gain}

\begin{figure}[!t]
\centering
\includegraphics[width=0.70\columnwidth]{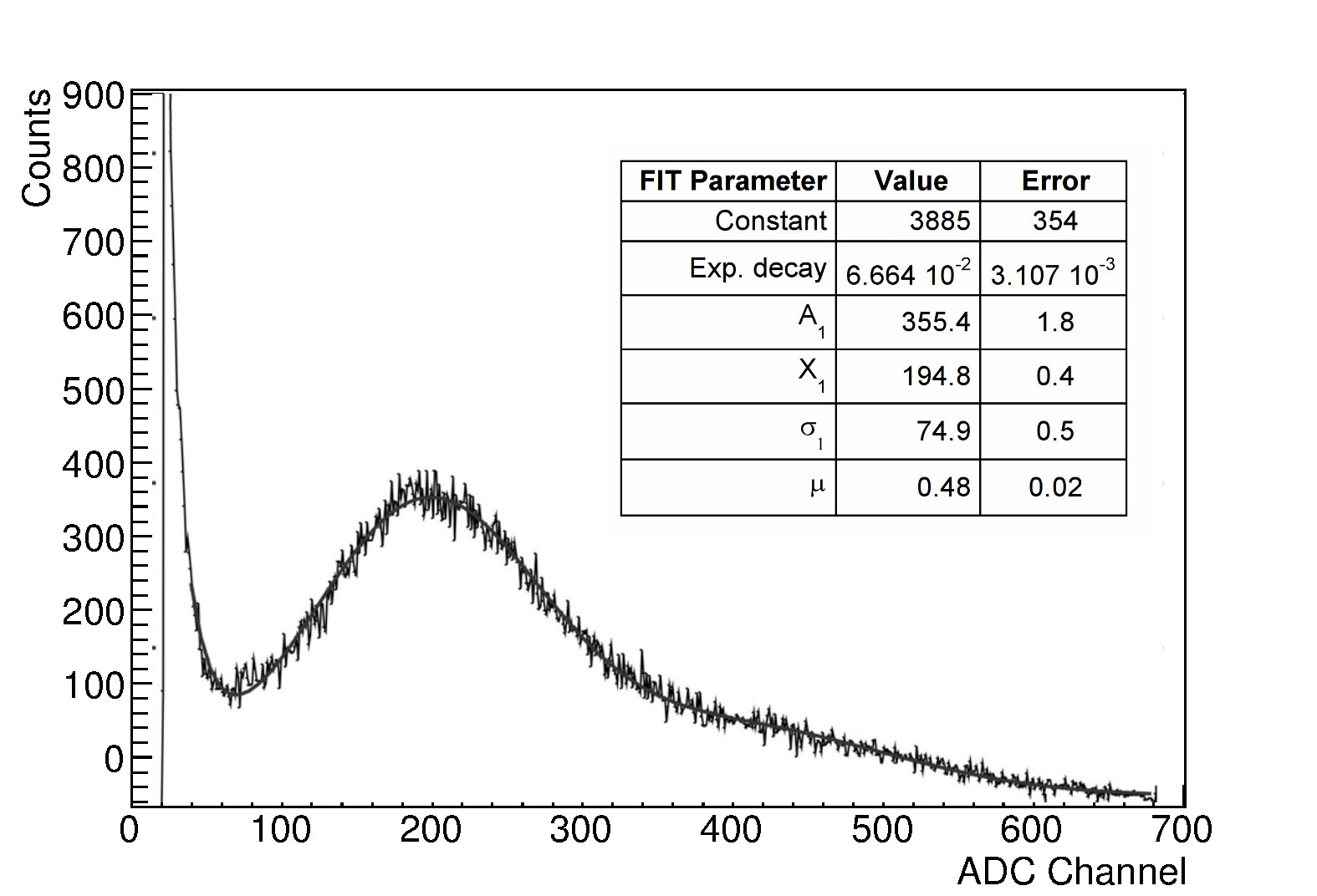}
\caption{Example of charge distribution for anode signals under single-photon illumination (SER).
The result of the fit, using the analytical expression described in the text, is also shown.}
\label{fig_ser}
\end{figure}

The response of a PMT to single photoelectrons (SER),
i.e., the charge distribution of the PMT pulses integrated
over the whole signal shape, is directly related to PMT
gain processes. SER studies were carried out by
measuring the charge distribution of the PMT pulses
induced by single-electron excitation at different
power supply values. An example of charge distribution is
shown in figure~\ref{fig_ser}.

Charge distributions are characterized by peak
profiles which feature a good charge
resolution. Spectra were fitted by means of an
analytical expression which consists of an
exponential distribution, which
takes into account PMT and
electronic noise, and 3
Gaussian functions which consider the
response of the PMT to different photoelectrons.
The following parameter constraints were used:\\
\centerline{
$X_n= nX_1$ \,\, $\sigma_n=\sigma_1\sqrt{n}$ \,\, $A_n = \frac{\mu}{n} A_{n-1}$}
where $X_n$ is the position of the $n^{th}$ Gaussian curve with
$\sigma_n$ width and $A_n$ amplitude. 
The value of the $\mu$ parameter,
resulting from the fit, represents the mean 
of the Poisson distribution of detected photoelectrons.

\begin{figure}[!t]
\centering
\includegraphics[width=0.7\columnwidth]{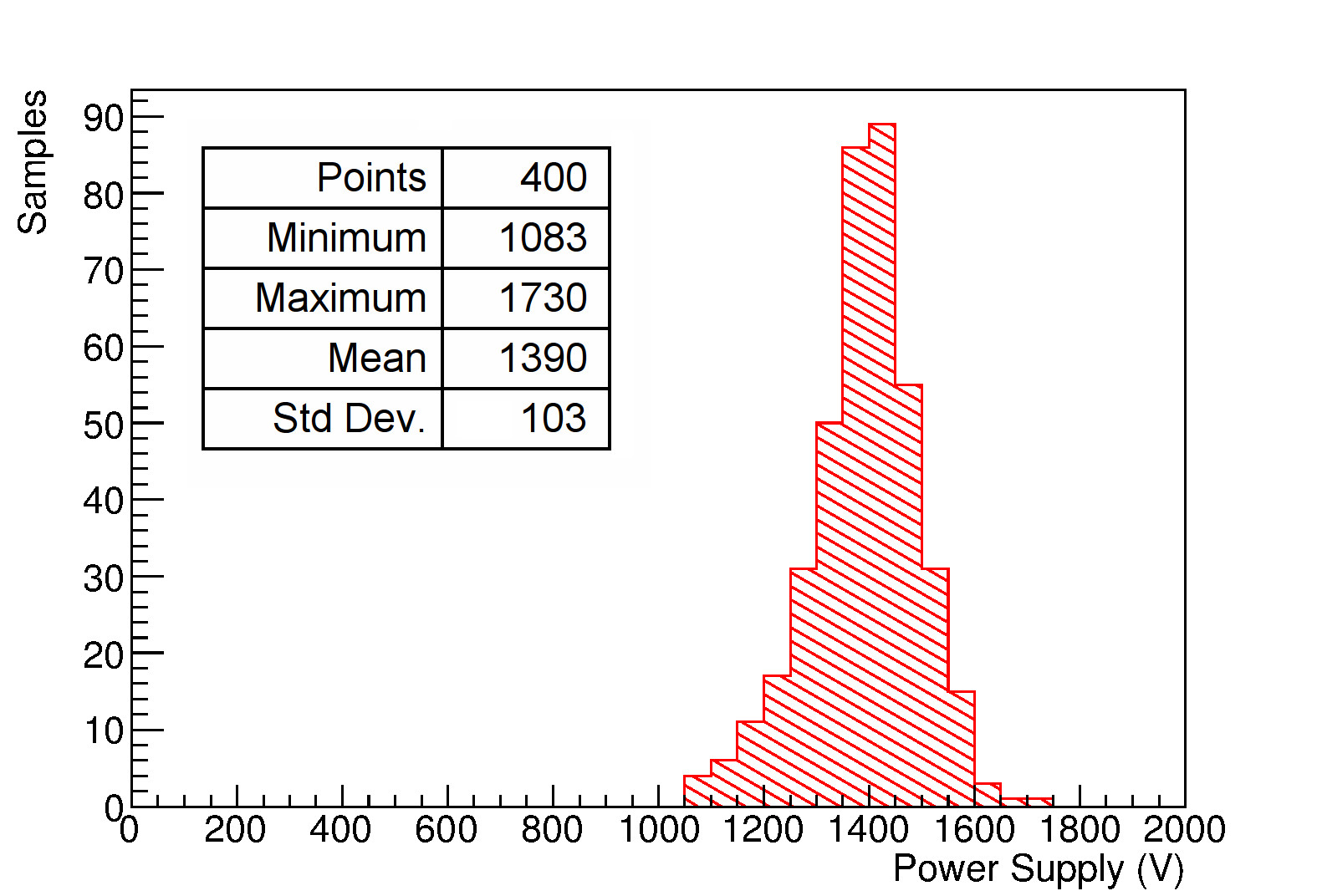}
\caption{Distribution of voltages to attain a nominal
gain $G = 10^7$
for the whole set of 400
PMTs operating at room temperature.}
\label{fig_HV_warm}
\end{figure}

The position $X_1$ of the first peak allows the evaluation of the gain $G$.
The gain dependence on the power supply applied to the PMT was measured by
changing the voltage setting in a range of values producing PMT gains
from about $G= 10^6$ to about $G=5\times 10^7$. 
The PMT nominal voltage 
was defined as the power supply value
to attain a gain $G= 10^7$. The nominal voltage distribution
for the whole set of 400 units operating at room temperature is shown in figure~\ref{fig_HV_warm}.
The distribution is characterized by a quite narrow spread with mean value set at $1390$~V
and $\sigma \approx 100$~V.


\begin{figure}[!t]
\centering
\includegraphics[width=0.7\columnwidth]{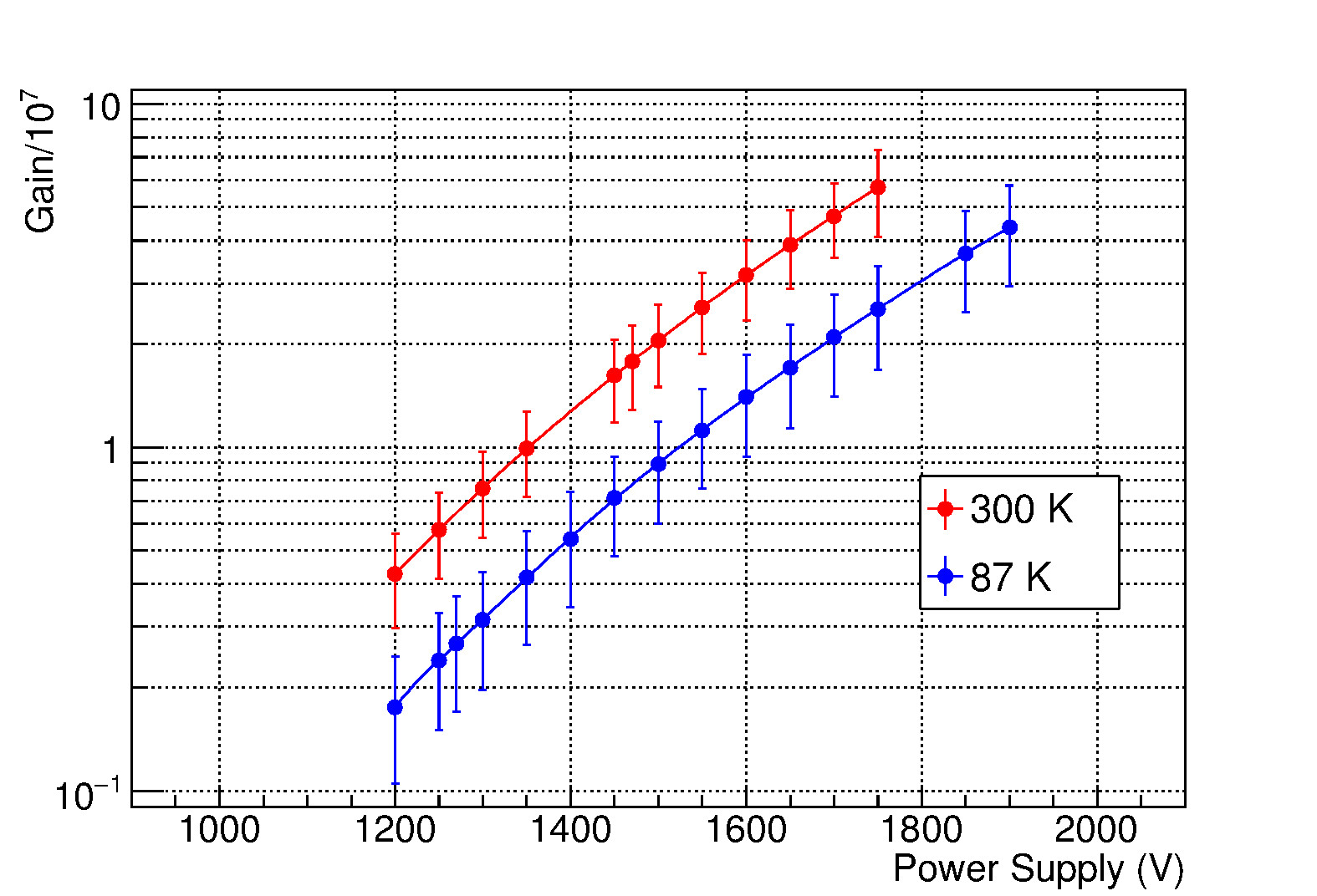}
\caption{Example of signal amplitude (Gain) as a function of the power supply voltage at room and
at cryogenic temperatures for a single PMT. For each measurement the peak position (dots) and 
the SER distribution width (vertical bars) coming from the fit are shown.}
\label{fig_gain}
\end{figure}

As an example, the results obtained at room and at cryogenic
temperatures, using a single PMT, are plotted in figure~\ref{fig_gain}. 
The signal amplitude is well represented by a power law
behavior as a function of the power supply voltage for both
temperatures. 
A remarkable reduction of the gain is evident at cryogenic
temperature with respect to results of measurements at room temperature.

\begin{figure}[!t]
\centering
\includegraphics[width=0.7\columnwidth]{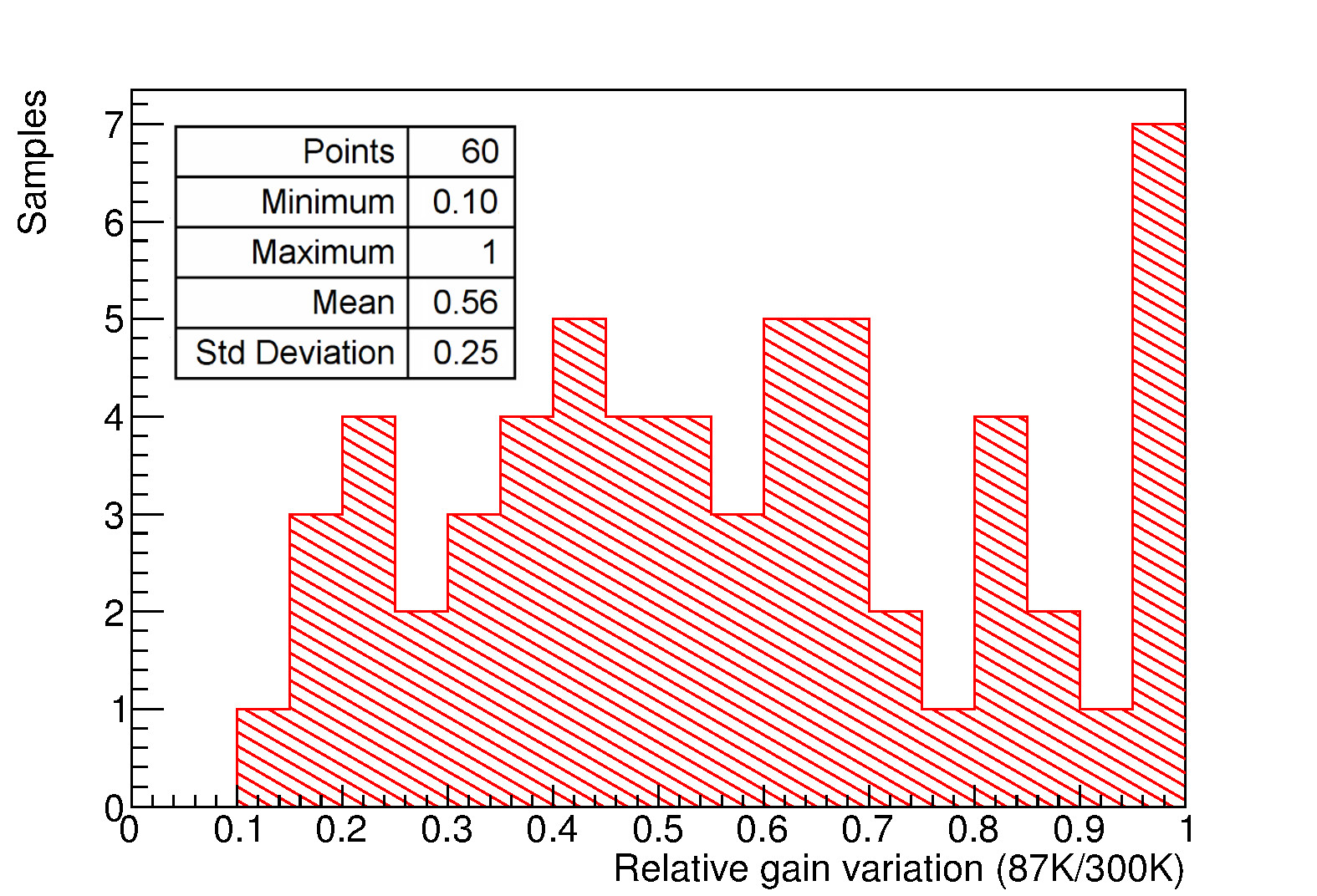}
\caption{Relative gain variation between room and cryogenic temperatures
of the 60 tested PMTs. 
Results are referred to a power supply consistent with a gain $G=10^7$ at room temperature.}
\label{fig_var}
\end{figure}

\begin{figure}[!t]
\centering
\includegraphics[width=0.7\columnwidth]{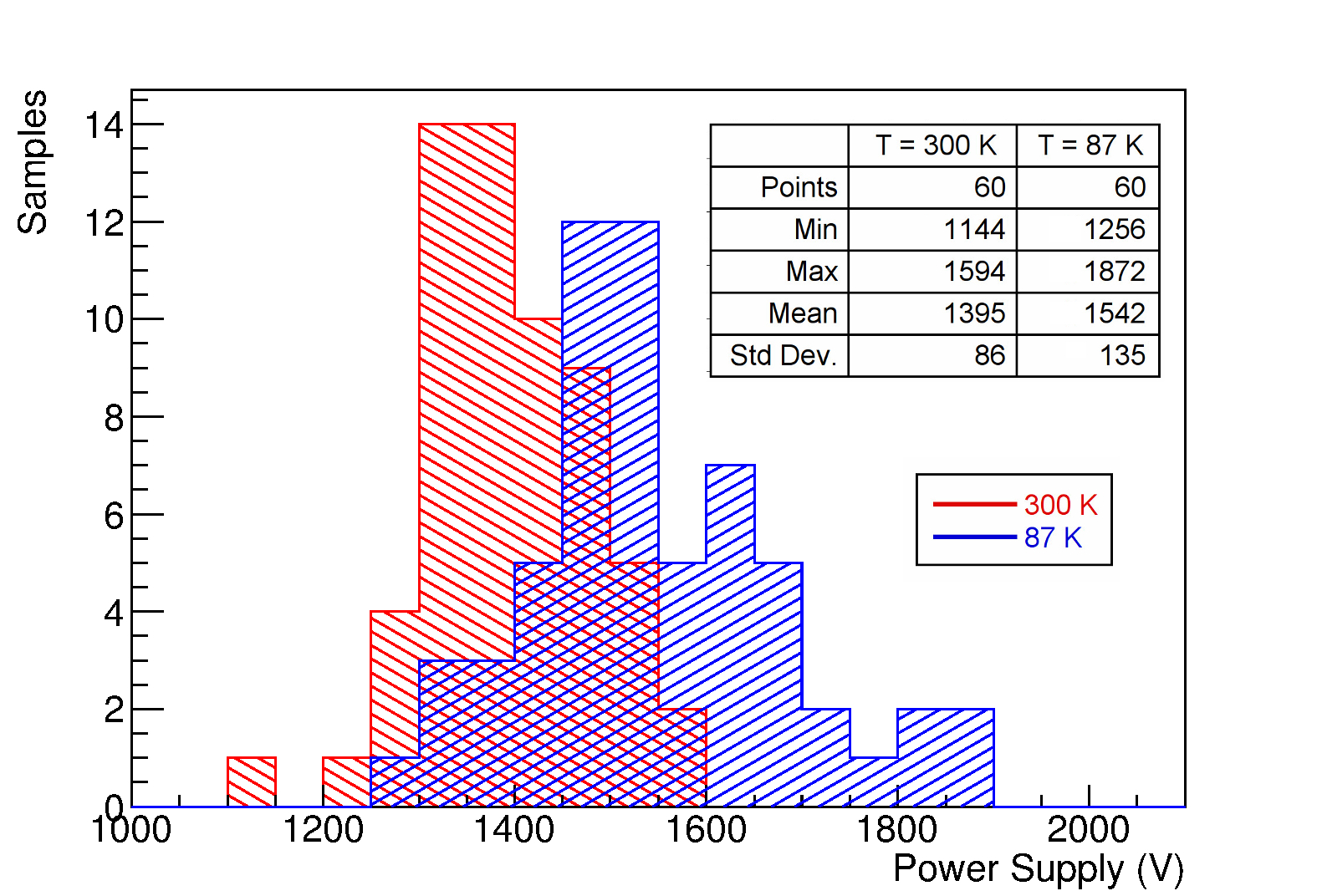}
\caption{Distribution of voltages  to
attain a gain $G= 10^7$  both for room
and cryogenic temperatures} 
\label{fig_HV}
\end{figure}

The distribution of the relative gain variation between room and
cryogenic temperature,
for the whole set of 60 devices tested in both conditions, is plotted in figure~\ref{fig_var}.
Almost all the tested PMTs suffered a gain reduction 
lasting three days in LAr bath. In this condition,
starting from a gain $G= 10^7$,
the PMTs showed wide gain decreases
down to 10\% of nominal values. During the time of observation,
the gain values at cryogenic temperature exhibited 
sudden drops before reaching a stable behavior, 
affecting the range of variation.
Anyway, an increase of less
than 150~V of the power supply voltage was on average
enough to recover the original gain factors, as shown in figure~\ref{fig_HV}.

The resolution of the SER peak, or relative variance to the peak,
defined as the ratio $\sqrt{s} = \sigma_1/X_1$, where $X_1$ is the SER peak position
and $\sigma_1$ is the width resulting from the fit, 
was determined for each PMT.
The mean results obtained with the
PMTs under test operating at a multiplier gain $G =10^7$ are:\\
\centerline{
\begin{tabular}{ll}
T = 300 K & $\sqrt{s} = 0.35 \pm 0.03$ (Std. Deviation)\\
T =  87 K & $\sqrt{s} = 0.46 \pm 0.11$ (Std. Deviation)\\
\end{tabular}}
These values demonstrate the
good performance in term of SER charge resolution
at both temperature environments, with a slight worsening
at 87~K.

\subsection{Dark counts and noise}

The response of a PMT in absence of light is represented by dark pulses and noise
mainly due to
ohmic leakage between electrodes on the glass and insulating surfaces of
the tubes, thermo-ionic emission of single electrons from the cathode, field emission of
electrons induced by local high electric fields inside the PMT. 

\begin{figure}[!t]
\centering
\includegraphics[width=0.7\columnwidth]{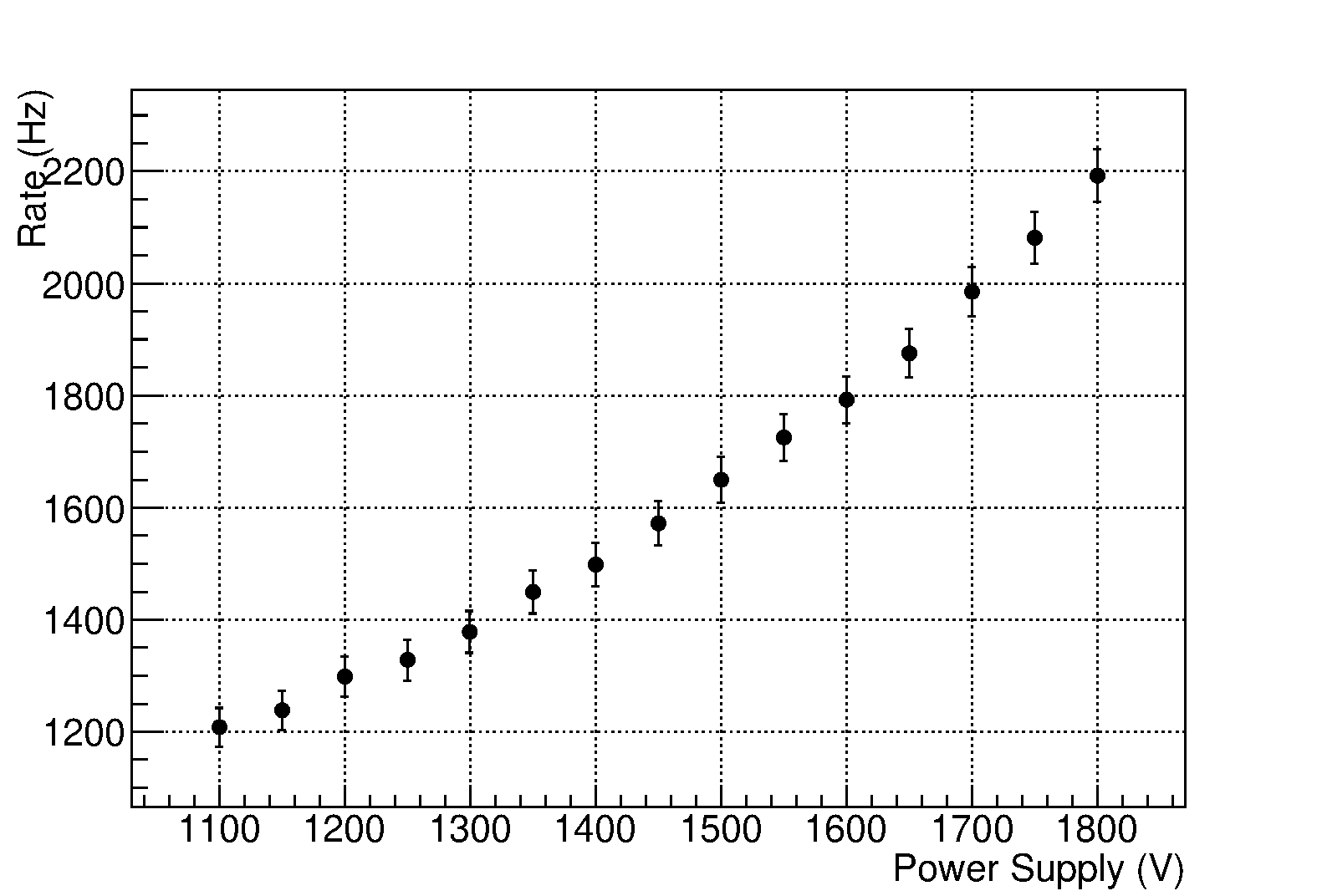}
\caption{Example of PMT dark rate as a function of the power supplier voltage level. 
The observed counting spreads are represented with vertical bars.}
\label{fig_darkhv}
\end{figure}

\begin{figure}[!t]
\centering
\includegraphics[width=0.7\columnwidth]{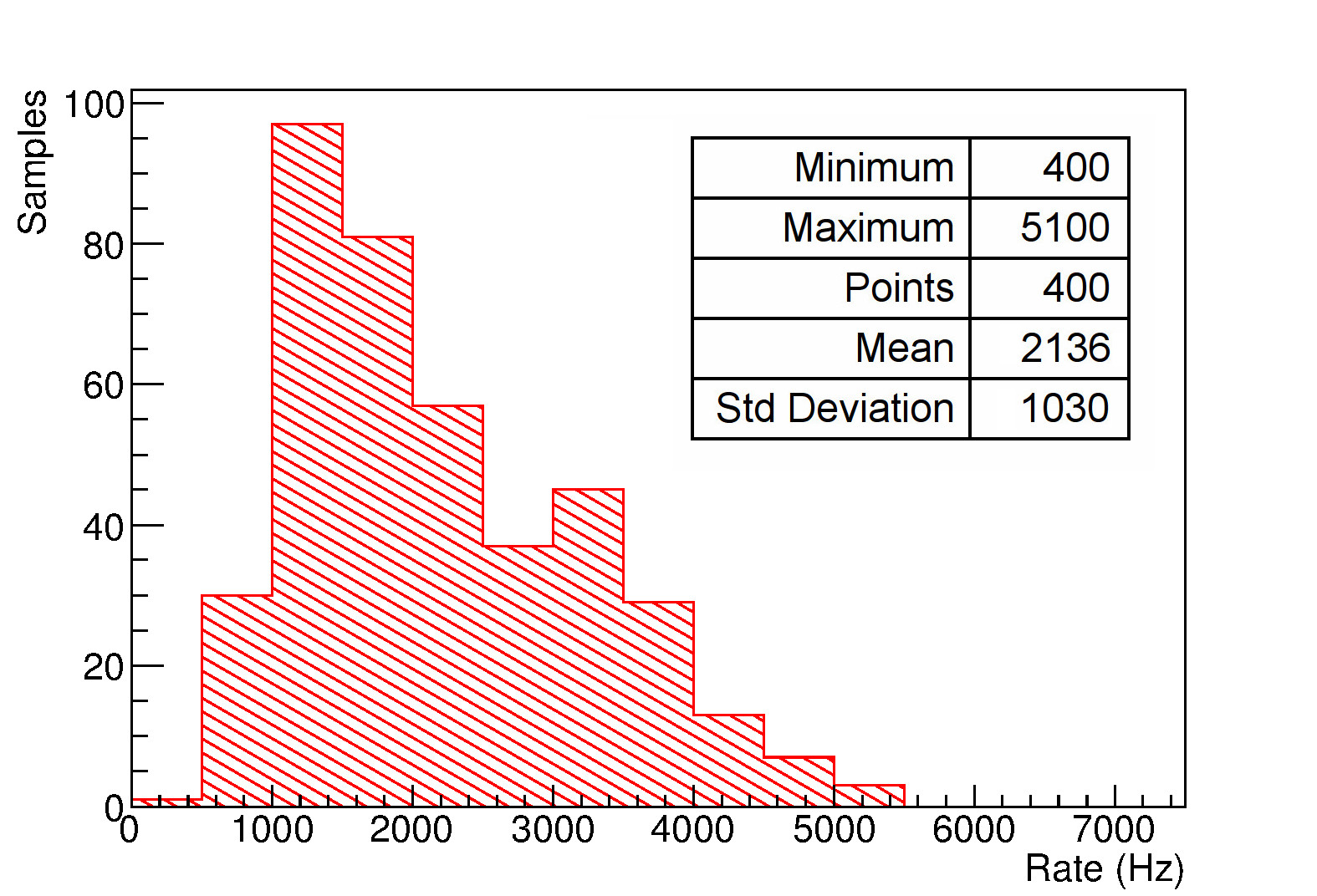}
\caption{Dark count rate at room temperature for the 400 tested PMTs.}
\label{fig_drate}
\end{figure}

The dark count rate was evaluated by measuring the 
rate of the PMT pulses, operating in darkness condition, 
with the discrimination threshold level set to the minimum value between the
pedestal and the maximum of the SER spectrum.
An example of dark count rate as a function of the power supply voltage is shown 
in figure~\ref{fig_darkhv}, while the distribution obtained
at room temperature and at a gain $G=10^7$ for the 400 tested PMTs is shown in figure~\ref{fig_drate}.

\begin{figure}[!t]
\centering
\includegraphics[width=0.7\columnwidth]{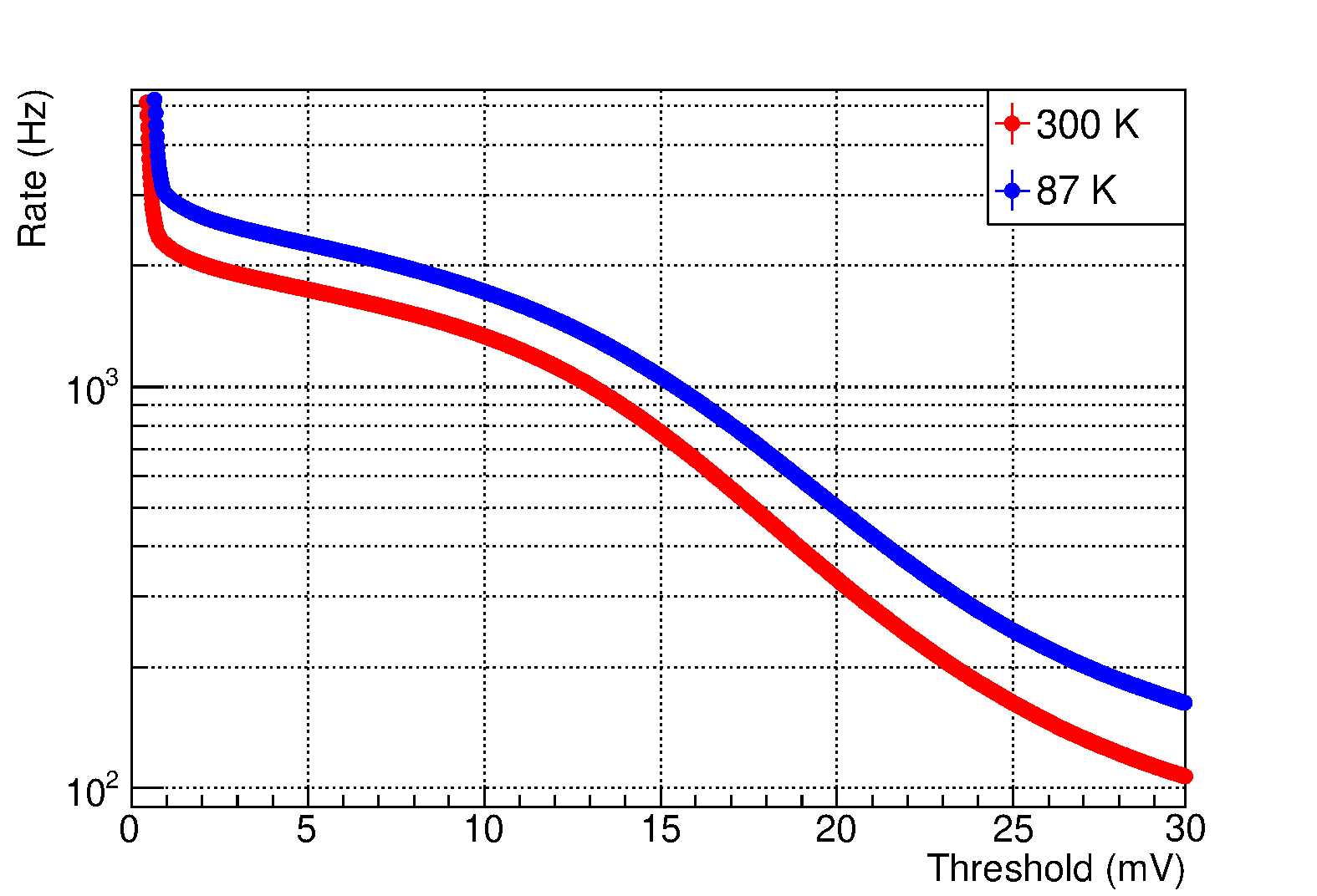}
\caption{
PMT dark count rate as a function of the discrimination
threshold at room and at cryogenic temperatures.}
\label{fig_dark}
\end{figure}

The increase of dark rate when operating a PMT at cryogenic temperature is a well-known effect, 
referred as Non-Thermal Dark Rate. It was observed and described by many
authors, as reported in \cite{Dark} and in references therein. The precise evaluation of the dark count rate 
with a PMT directly immersed in liquid argon is not an easy task, due to the
presence of Cherenkov radiation and scintillation produced by residual radioactive
contaminants, such as $^{39}$Ar. In order to prevent these effects, 
the measurements were carried out using a stainless steel
chamber designed to house the PMT under
test in vacuum and dark environment. The chamber was placed in a dewar that
was filled with liquid argon during the cryogenic measurement.
This test, which takes a quite long time for the thermalization,
was carried out only on 5 PMT samples.
An example of resulting
rates, as a function of the discrimination threshold and for
the two different operating temperatures is shown in figure~\ref{fig_dark}.
The curves were acquired 
at a multiplier gain $G\approx  10^7$.
The spectrum profile is characterized by a hump
centered in the region around one photoelectron,
caused mainly by the cathode dark noise. An increase of
the counting rate at cryogenic temperature, up to a factor 2, was experienced
for all the tested PMTs.

\begin{figure}[!t]
\centering
\includegraphics[width=0.7\columnwidth]{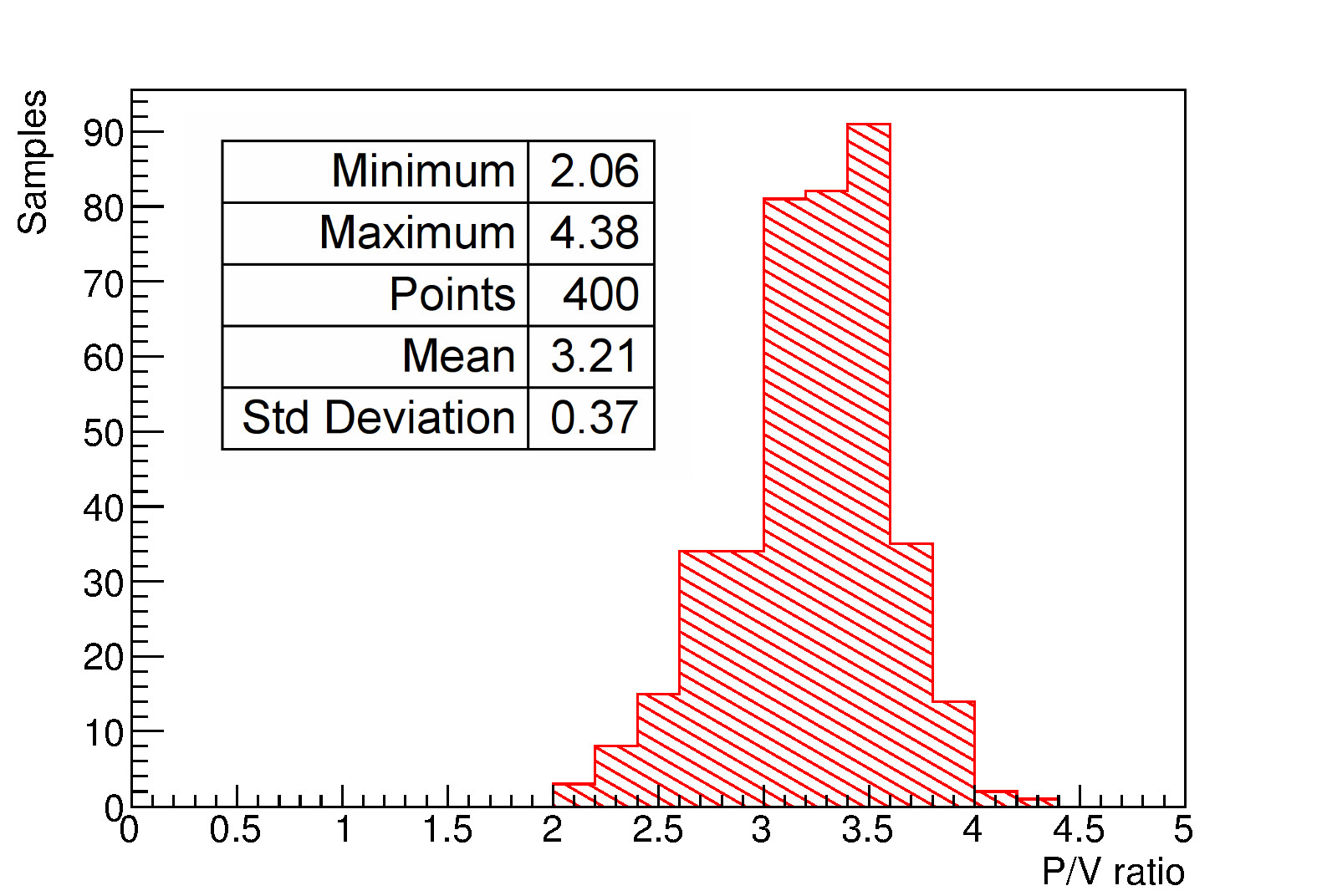}
\caption{Distribution of the $P/V$ ratio at room temperature for the
400 tested PMTs. Results are referred to a power supplier voltage
level consistent
with a gain $G=10^7$.}
\label{fig_PVall}
\end{figure}

\begin{figure}[!t]
\centering
\includegraphics[width=0.7\columnwidth]{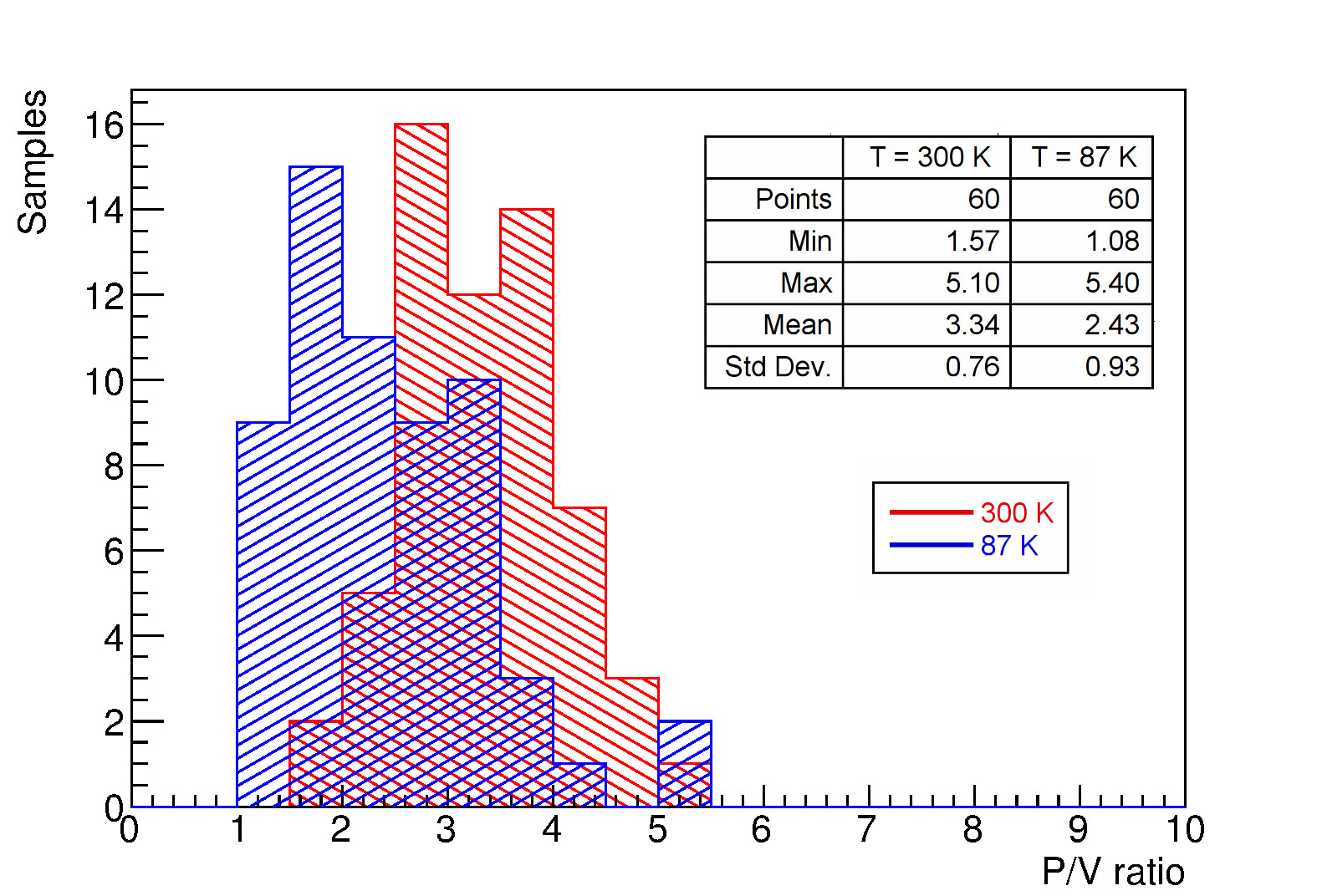}
\caption{Comparison of the distribution of the
$P/V$ ratio at room (300 K) and at cryogenic (87 K) temperatures for
60 tested PMTs. Results are referred to a power supply consistent
with a gain $G=10^7$ at both temperatures.}
\label{fig_pv}
\end{figure}

The peak-to-valley $P/V$ ratio, defined 
as the SER peak value divided by the minimum
value to the left of the peak, is a typical indicator of the PMT capability to
distinguish actual photoelectron signals from its intrinsic noise.
The $P/V$ distribution
at room temperature for the 400 tested PMTs operating at $G =10^7$
is shown in figure~\ref{fig_PVall}.
The mean value, resulting from the SER distribution fitting,
exceeds $P/V = 3:1$, marking very good performances of this
PMT model in terms of signal/noise ratio.
The variation of the $P/V$ ratio
between room and cryogenic temperature, for the 60 PMTs tested in LAr bath,
is shown in figure~\ref{fig_pv}. 
A decreases of less than 30\% at cryogenic temperature, mainly due to
an increase of the PMT intrinsic noise at that temperature, is observed.

\subsection{Photocathode uniformity}

The evaluation of the photocathode uniformity 
required dedicated tests and was carried out 
for the 20 pre-series samples at room temperature, only.
The PMTs were operated as photodiodes. To this
purpose all the dynodes and the anode were shorted and grounded
to collect all the photoelectrons emitted
by the cathode which was kept at a negative voltage of about
300~V. 

A  collimated laser diode (Thorlabs CPS532,
$\lambda = 532$~nm, 4.5~mW) was used as a continuous light source
producing  cathodic currents of the order of 10~$\mu$A, 
directly measured by a picoammeter.
In order to monitor the illumination intensity and its stability,
the light was split toward a photometer and the PMT
which was illuminated by
means of an optical fiber; a proper support was used to maintain the fiber
in a fixed orientation, normal to the PMT window, while allowing to move
it in various positions on the window itself. 

\begin{figure}[!t]
\centering
\includegraphics[width=0.7\columnwidth]{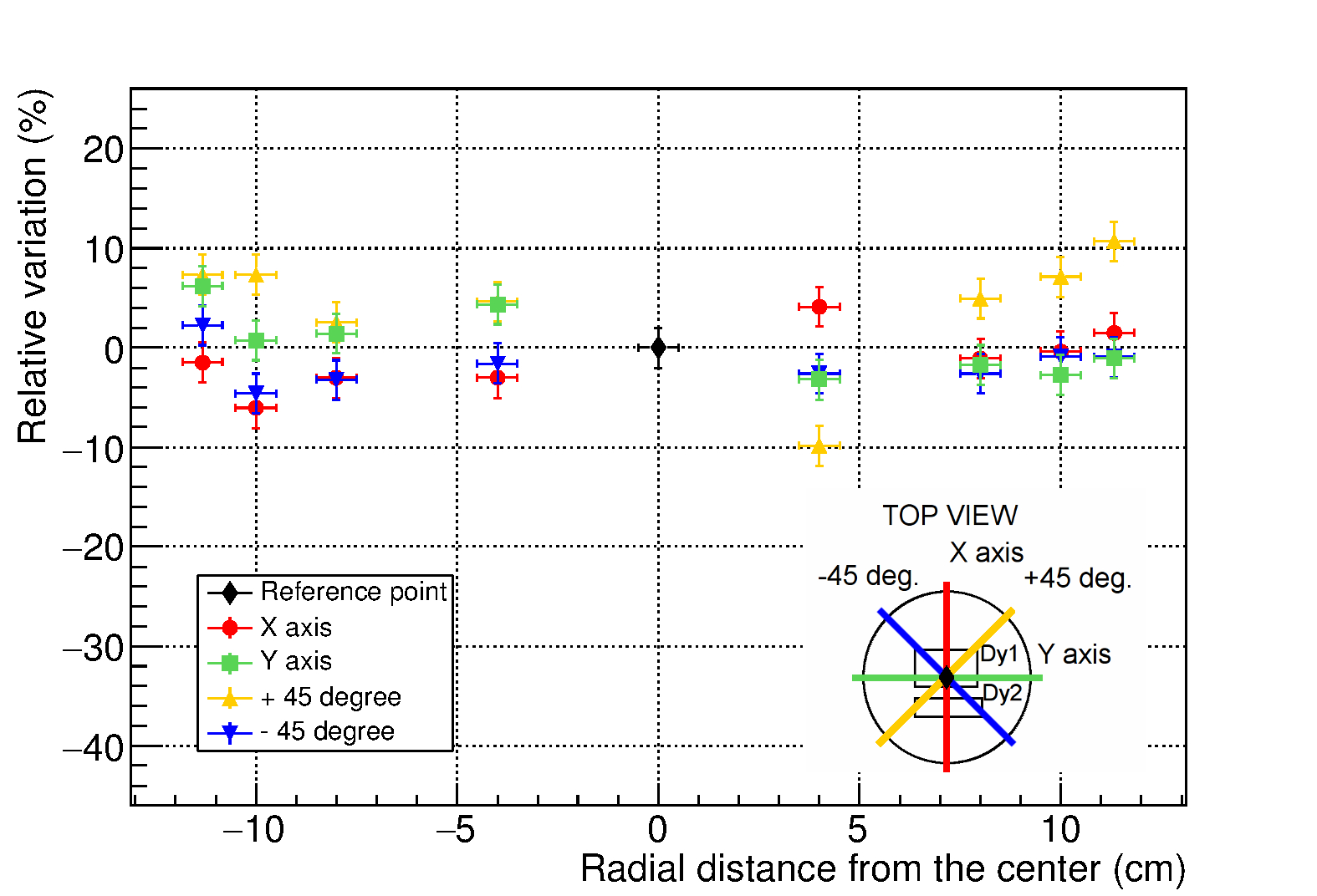}
\caption{Variation of the photocathode response as a
function of the radial distance from the window center along 4 axes oriented at
$45^\circ$ each other.}
\label{fig_uni}
\end{figure}

Figure~\ref{fig_uni} shows, as example, the variation of the a photocathode response as a
function of the radial distance from the window center along four different axes oriented at
$45^\circ$ each other. 
An increase of the photocathode signal
approaching the window edges is recognized. This behavior was observed for all the 20 tested PMTs and it was
confirmed by measurement performed by the producer.

\begin{figure}[!t]
\centering
\includegraphics[width=0.70\columnwidth]{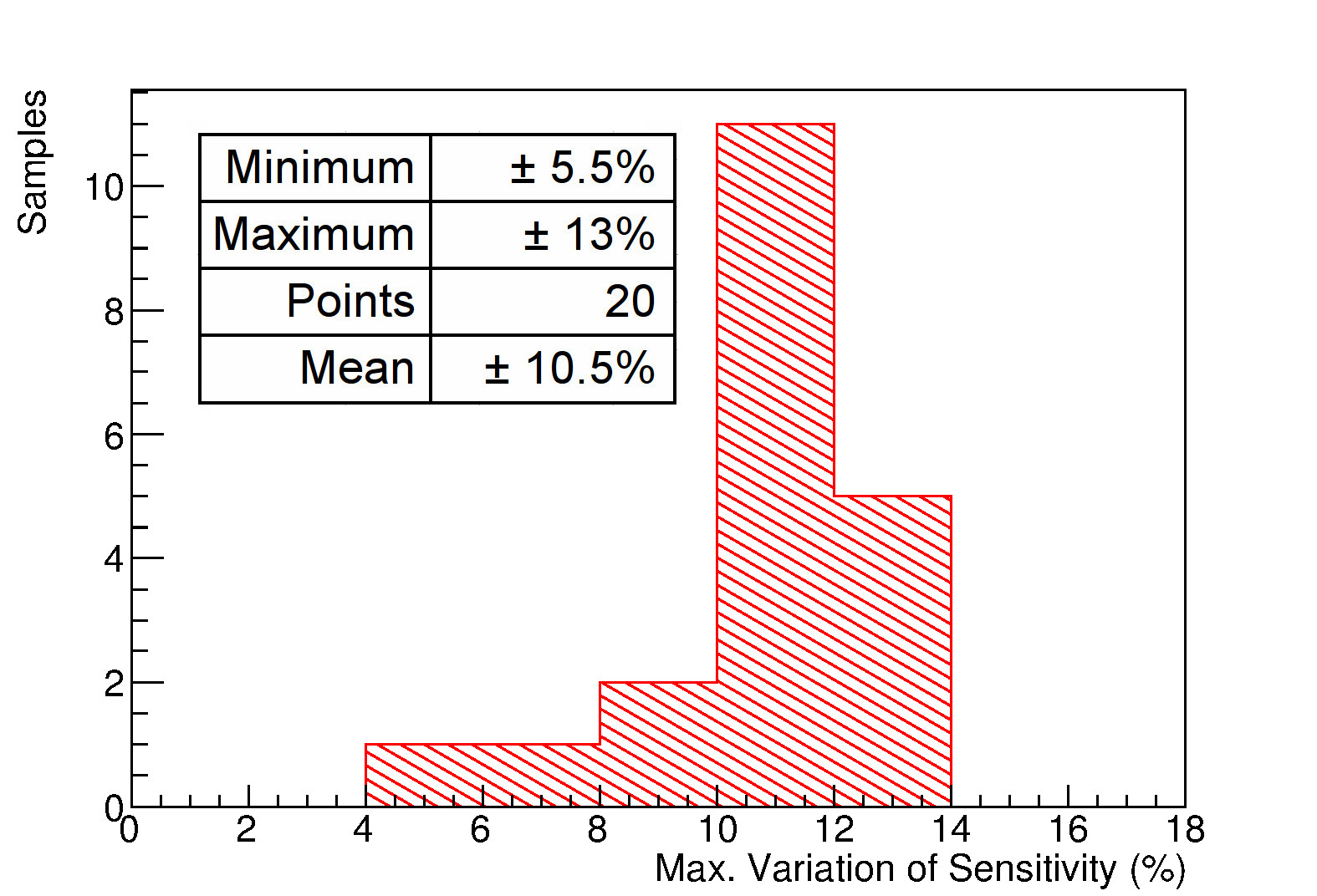}
\caption{Distribution of the maximum variation with respect to the mean value of the PMT response for the 20
PMT samples.}
\label{fig_univar}
\end{figure}

An approximate comparison among different samples was made by averaging, for each PMT, the 
response data over the sensitive surface and evaluating the maximum relative variation with respect to the mean value. The resulting distribution is shown in figure~\ref{fig_univar}. 
All the 20 PMTs show a good uniform response, within $\pm 13 \%$ in agreement with the
values indicated by the manufacturer.

\subsection{Discharge and light emission}

Some authors report that, at high voltage, in some PMT models it is possible to
observe the presence of discharges which produce photons polluting the output
signals (see for example~\cite{Akimov}).
This phenomenon is not yet well understood but it is characterized by the presence of both
large light pulses (``flashes'') and microscopic processes at the SER level, also in
cryogenic environment.
 
A fast check of a possible presence of light emission was 
carried out during the characterization at cryogenic temperature
of 60 PMTs.
As described in Section~{\ref{sec_equip}}, measurements were performed with the
simultaneous immersion of 10 PMTs in LAr. The count rate of each PMT was recorded
keeping the sample under test permanently powered ON and turning, in sequence,
all the other 9 specimens ON and OFF.

Variations of the count rate between the two different operating conditions
were always within the statistical errors. In addition, the presence of signal
coincidences beyond the expected stochastic values were excluded in the data,
minimizing the possible presence of 
flashes or other PMT light emission phenomena that could affect the correct operation of the 
tested PMTs.

\section{Conclusions}

The updated ICARUS T600 light detection system consists of 360 Hamamatsu R5912-MOD PMTs
directly operating in liquid argon to reveal scintillation photons produced by
ionizing particles. A total of 400 PMTs were characterized at CERN at room temperature 
and 60 of them at cryogenic temperature
to evaluate any parameter variation which could affect the scintillation light detection.

The main obtained results can be summarized as:
\begin{itemize}
\item SER and signal shape parameters are 
in good agreement with the producer nominal values, both at room and
at cryogenic temperature;
\item The resulting gain reduction at cryogenic temperature,
down to 10\% of the nominal value, can be on average compensated by an increase of less than 150~V of
the power supply voltage;
\item Dark counts and noise are in agreement with the producer nominal values at room temperature. The
observed worsening at cryogenic temperature of dark counts and P/V ratio is still compatible
with the correct operation of this PMT model in LAr bath;
\item Good uniformity of the response of the cathode was observed all over the sensitive window;
\item The presence of lighting phenomena were not observed;
\item No mechanical damages, implosions or cracking problems were observed neither during cooling/heating 
phases nor operating at cryogenic temperature.
\end{itemize}

All the tested PMTs were rated as compliant with the requirements of ICARUS T600
and a subset of 360 specimens were selected for the final installation in the detector.

\section*{Acknowledgement}
This work was funded by INFN in the framework of CERN WA104/NP01 program finalized to the
overhauling of ICARUS detector in view of its operation on SBN at Fermilab.
The staff of Hamamatsu Photonics K.K. Japan, Italy and Switzerland is acknowledged
for the fruitful technical collaboration and discussion.

\appendix
\section{Custom PMT base layout}

\label{base}

\begin{figure}[!t]
\centering
\includegraphics[width=0.45\columnwidth]{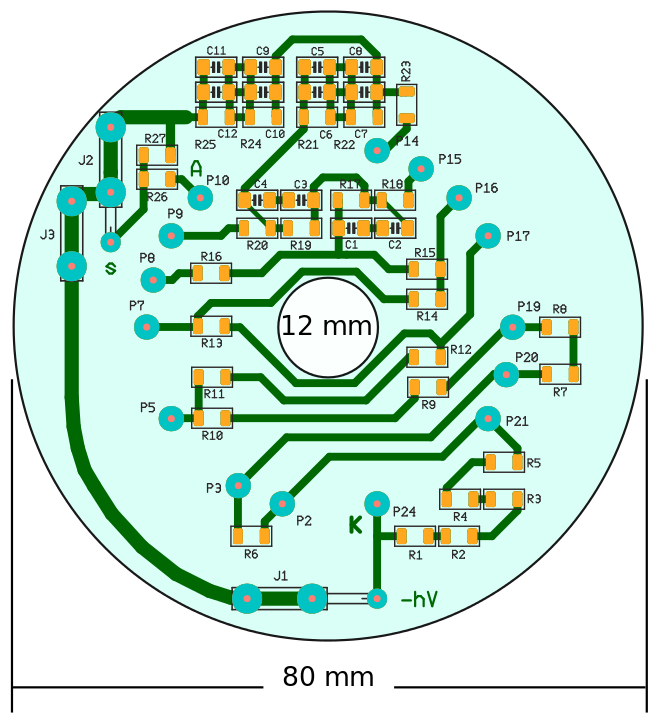}
\includegraphics[width=0.45\columnwidth]{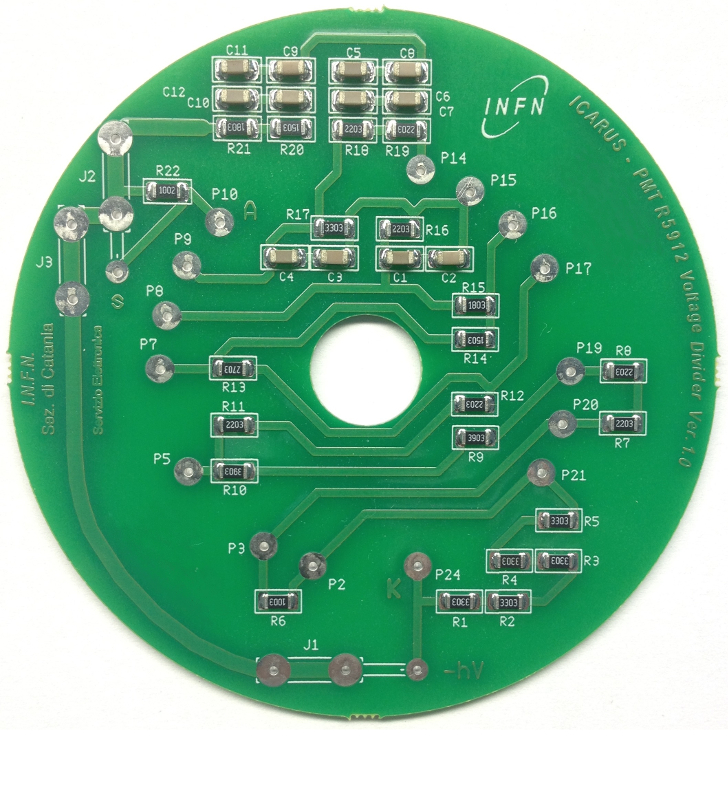}
\caption{PMT base circuit, design (left) and picture (right).}
\label{fig_base}
\end{figure}

\begin{figure}[!t]
\centering
\includegraphics[width=1\columnwidth]{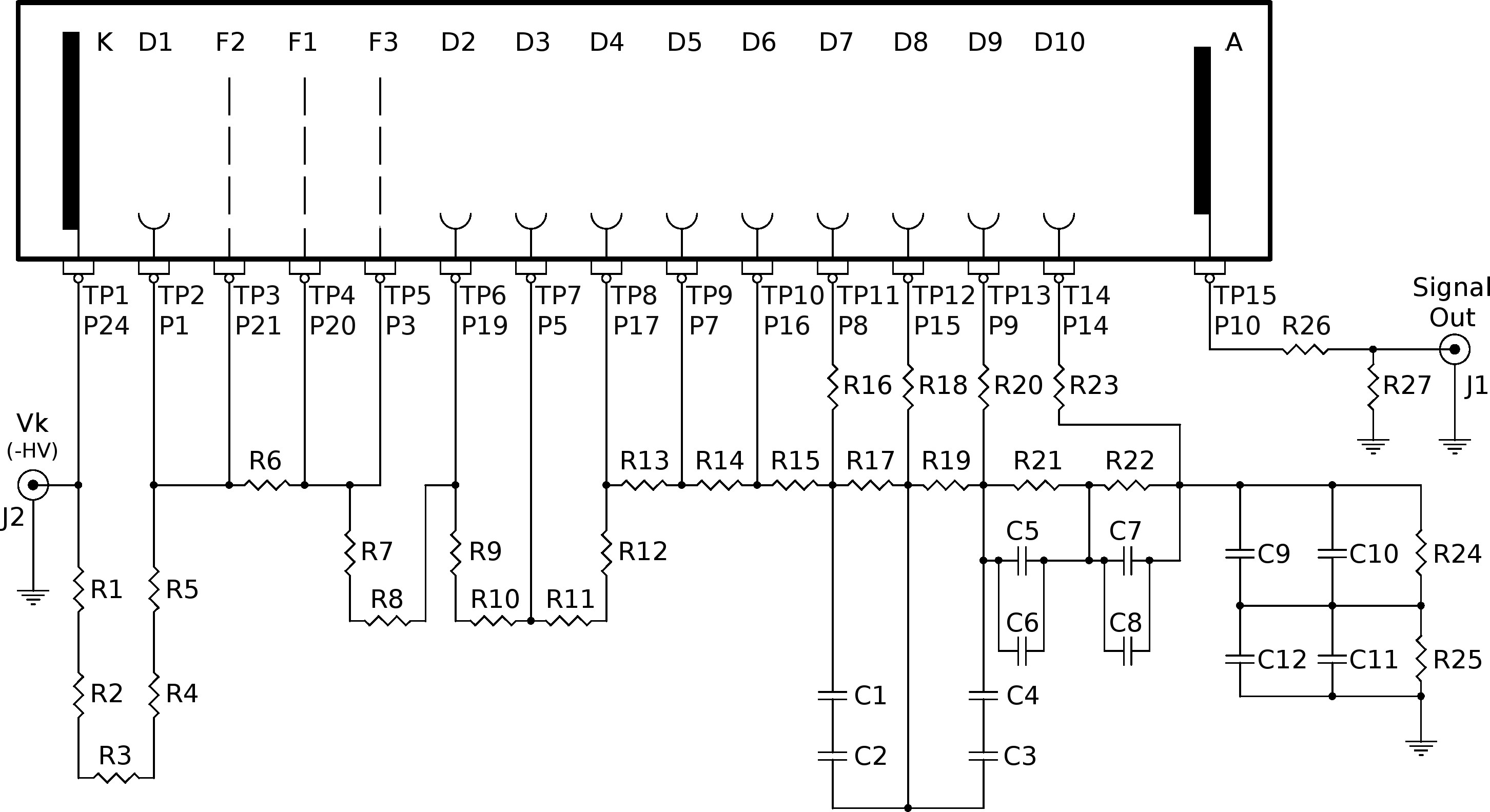}
\caption{PMT voltage divider}
\label{fig_divider}
\end{figure}

The adopted PMT base 
is shown in figure~\ref{fig_base} and the 
main circuit properties are presentet in table~\ref{table2}.
The resistive divider is outlined in figure~\ref{fig_divider}.
The reference voltage distribution
ratio is the standard recommend by Hamamatsu (table~\ref{table3}).
A particular care was devoted to the choice of damping resistors to improve
the PMT time response. The $100 \Omega$ resistors, mounted in series to each
of the last three dynodes, perform a good reduction of the ringing in the output waveform.

The voltage divider chain is entirely passive, and is fabricated
with SMD resistors and capacitors 1206 package (table~\ref{table4}), all tested in LAr bath
before the mounting.
The increase of the resistor value at cryogenic temperature is $\approx 3\%$
for all the components, leaving unaffected the voltage distribution
ratio of the divider. The nominal value of all the capacitors, with C0G dielectric, 
is maintained  even at cryogenic temperature.


\vspace{5cm}

\begin{table}[!t]
\centering
\caption{PMT base circuit characteristics}
\label{table2}
\begin{tabular}{ll}
\hline
Board material & FR-4 \\
Board thickness & 1.6~mm \\
Thrugh holes type & plated\\
Hole size (before plating) & 0,9~mm\\
Copper thickness & $35 \mu$m \\
Soldermask & green \\
Silkscreen & white \\
\hline
\end{tabular}
\end{table}


\begin{table}[!t]
\renewcommand{\arraystretch}{1}
\caption{Voltage distribution ratio}
\label{table3}
\centering
\begin{tabular}{cc}
\hline
Electrodes & Ratio \\
\hline
K-D1   & 16.8 \\
D1-F2  &  0 \\
F2-F1  & 0.6 \\
F1-F3  &  0 \\
F3-D2  & 3.4 \\
D2-D3  & 5 \\
D3-D4  & 3.33 \\
D4-D5  & 1.67 \\
D5-D6  &  1 \\
D6-D7  & 1.2 \\
D7-D8  & 1.5 \\
D8-D9  & 2.2 \\
D9-D10 & 3 \\
D10-A  & 3.4 \\
\hline
\end{tabular}
\end{table}

\begin{table}[!t]
\renewcommand{\arraystretch}{1}
\caption{Components}
\label{table4}
\centering
\begin{tabular}{ll}
\hline
C1 C2 C3 C4 C5 C6 C7 C8 C9 C10 C11 C12 & 22~nF$\pm 5\%$\\
R1 R2 R3 R4 R5 R19                     & 330~k$\Omega \pm 1\%$\\
R6                                     & 100~k$\Omega \pm 1\%$\\
R7 R8 R11 R12 R17 R21 R22              & 220~k$\Omega \pm 1\%$\\
R9 R10                                 & 390~k$\Omega \pm 1\%$\\
R13                                    & 270~k$\Omega \pm 1\%$\\
R14 R24                                & 150~k$\Omega \pm 1\%$\\
R15 R25                                & 180~k$\Omega \pm 1\%$\\
R18 R20 R23                            & 100~$\Omega \pm 1\%$ \\
R26                                    & 51~$\Omega \pm 1\%$ \\
R16                                    & 0~$\Omega $ \\
\hline
All components are SMD, 1206 package, 200~V rated.& \\
Capacitors are C0G dielectric.& \\
\end{tabular}
\end{table}

$\,$

\end{document}